\documentclass[12pt]{article}
\usepackage{epsfig}
\input epsf.sty
\usepackage{pstricks}
\newcommand{\be}{\begin{equation}}
\newcommand{\ee}{\end{equation}}
\newcommand{\ba}{\begin{eqnarray}}
\newcommand{\ea}{\end{eqnarray}}
\newcommand{\tr}{{\rm Tr\,}}
\newcommand{\ii}{{\rm i}}

\newcommand{\ex}{{\rm e}}
\newcommand{\nn}{\nonumber}

\newcommand{\bfx}{{\bf x}}

\newcommand{\bmu}{\bar{\mu}}

\newcommand{\op}{{\cal O}}
\newcommand{\eq}{Eq.~}
\newcommand{\eqs}{Eqs.~}
\newcommand{\fig}{Fig.~}
\newcommand{\RR}{{\rm I\kern -.2em  R}}
\def\lsi{\raise0.3ex\hbox{$<$\kern-0.75em\raise-1.1ex\hbox{$\sim$}}}
\def\gsi{\raise0.3ex\hbox{$>$\kern-0.75em\raise-1.1ex\hbox{$\sim$}}}
\newcommand{\lsim}{\mathop{\lsi}}

\def\none               {\multicolumn{2}{c|}{--}}

\begin{document}

\begin{titlepage}
\begin{flushright}
CERN-TH/2003-154 \\
MIT-CTP-3395
\end{flushright}
\begin{centering}
\vfill

{\bf \Large The QCD Phase Diagram for Three Degenerate Flavors and Small Baryon Density}

\vspace{0.8cm}

Philippe~de Forcrand$^{\rm a,b}$ and
Owe~Philipsen$^{\rm c}$

\vspace{0.3cm}

{\em $^{\rm a}$
Institut f\"ur Theoretische Physik,
ETH Z\"urich,
CH-8093 Z\"urich, 
Switzerland\\}
{\em $^{\rm b}$
Theory Division, CERN, CH-1211 Geneva 23,
Switzerland\\}
{\em $^{\rm c}$
Center for Theoretical Physics, Massachusetts Institute of Technology,\\
Cambridge, MA 02139-4307, USA}

\vspace*{0.7cm}

\begin{abstract}\noindent
We present results for the phase diagram of three flavor QCD
for $\mu_B\lsi 500$ MeV.
Our simulations are performed with imaginary chemical potential $\mu_I$
for which the fermion determinant is positive. Physical observables
are then fitted by truncated Taylor series and continued to
real chemical potential. We map out the location of the critical
line $T_c(\mu_B)$ with an accuracy up to terms of order $(\mu_B/T)^6$.
We also give first results on a determination of the critical endpoint
of the transition and its quark mass dependence. Our results 
for the endpoint differ significantly
from those obtained by other methods, and we discuss possible reasons for this.
\end{abstract}

\end{centering}
\noindent
\vfill
\end{titlepage}

\section{Introduction}

The last two years have seen significant progress in simulating
QCD at small baryon densities. Standard Monte Carlo methods fail
in the presence of a non-vanishing chemical potential, 
for which the fermion determinant is complex and prohibits 
importance sampling with positive weights. While this problem remains
unsolved for QCD, there are presently three avenues to investigate the 
$(\mu,T)$-phase diagram,
as long as the quark chemical potential in units of temperature, $\mu/T$, is 
sufficiently small:
$(i)$ A two-dimensional generalization of the Glasgow method \cite{rew}
predicts a phase diagram with a first order phase transition at large $\mu$, terminating in 
a critical endpoint \cite{fk2}. 
$(ii)$ Taylor expanding the observables and the
reweighting factor leads to coefficients expressed in local operators and thus
permits the study of larger volumes \cite{hk}. 
$(iii)$ Simulations at imaginary chemical potential
are not limited in volume since the fermion determinant is positive. 
They allow for fits of the full observables
by truncated Taylor series, thus controlling the convergence of the latter, and subsequent
analytic continuation to real $\mu$ \cite{fp}.

The results for the location of the pseudo-critical line $T_c(\mu_B)$ are consistent
among all three approaches for baryon chemical potentials $\mu_B$ 
up to $\mu_B\lsi 500$ MeV ($\mu_B=3\mu$). This latter 
number gives the range of applicability of method $(iii)$, and hence the range over
which convergence of the Taylor series can be checked explicitly. 
A review comparing these methods in detail can be found in \cite{lap}. However,
the location of the endpoint of the phase transition 
has only been computed by one method and for one
set of quark masses \cite{fk2}. Since the numerical study of critical phenomena
is notoriously difficult even for $\mu=0$, it is important to cross-check this
result by another method.

In the present paper we investigate QCD with three degenerate quark flavors 
by lattice simulations with imaginary chemical potential. 
In this case we know reasonably well the chiral critical point $m_c(\mu=0)$, 
i.e.~the critical bare quark mass $m_c$ 
for which the deconfinement transition
changes from first order to crossover \cite{aoki,kls,nc}. This point
marks a second order phase transition
in the universality class of the 3d Ising model \cite{kls}. 
For the $(\mu,T)$-phase diagram this means that for $m<m_c$ the line of first order 
deconfinement/chiral phase transitions
extends all the way to the temperature axis at $\mu=0$, whereas 
for $m>m_c$ it terminates at a critical point $(T^*,\mu^*)$. The critical
chemical potential, $\mu^*(m)>0$, is expected to grow with the quark mass.
Inverting this relation yields the change in the 
critical bare quark mass with chemical potential, $m_c(\mu)$.
Our goal is to determine this function for chemical potentials $\mu_B\lsi 500$ MeV.

We begin by computing the quark mass and chemical potential dependence of the critical
line, $T_c(\mu_B,m)$. Our simulations are accurate enough 
to allow for a determination of the $(\mu_B^4)$ coefficient of its Taylor series. 
We then proceed to extract 
$m_c(\mu)$ by measuring the Binder cumulant 
of the chiral condensate. We find the $\mu$-dependence of the latter to be very weak,
proving a quantitative determination of $m_c(\mu)$ to be extremely difficult.
Just like the critical temperature, $m_c(\mu)$ is
an even function of the chemical potential with a Taylor expansion in $\mu^2$.
We are only able to determine the first non-trivial coefficient with an error of
40\%. However, this allows us to give a conservative upper bound on this
coefficient at the 90\% confidence level. This bound is in disagreement with
a preliminary result obtained by Taylor expanded reweighting \cite{alt1}.

After summarizing the imaginary chemical potential approach in Sec.~\ref{imag},
Sec.~\ref{gen} discusses general qualitative features and expectations 
about the three flavor phase diagram and how it can be 
determined by simulations at imaginary chemical potential.
Sec.~\ref{meth} introduces the Binder cumulant and its finite volume scaling
as our computational tool to determine the order of the phase transition
as a function of quark mass and chemical potential. 
Our numerical results are presented in Sec.~\ref{num}, followed by 
a discussion and comparison with other work in Sec.~\ref{dis}.
Finally, we give our conclusions.

\section{\label{imag} The QCD phase transition from imaginary $\mu$}

This section serves to fix the notation and summarize what is needed in the sequel.
For a detailed discussion of the formalism we refer to \cite{fp}.
The QCD grand canonical partition function
$Z(V,\mu,T)=\tr \left(\ex^{-(\hat{H}-\sum_f\mu \hat{Q_f})/T}\right)$, 
with the same chemical
potential $\mu$ for all flavors,
can be considered for {\em complex} chemical potential
$\mu = \mu_R + \ii \mu_I$. Two general symmetry properties can be used to
constrain the phase structure of the theory as a function of $\mu_I$:
$(i)$ $Z$ is an even function of $\mu$, $Z(\bmu)=Z(-\bmu)$,
where $\bar{\mu}=\mu/T$;
$(ii)$ A non-periodic gauge transformation, which rotates
the Polyakov loop by a center element but leaves $Z$ unchanged,
is equivalent to a shift in $\mu_I$~\cite{weiss}:
\begin{equation}
Z(\bmu_R,\bmu_I)=Z(\bmu_R,\bmu_I+2\pi/N).
\end{equation}
For QCD ($N=3$), these two properties lead to $Z(3)$ transitions at critical values of
the imaginary chemical potential,
$\bmu_I^c=\frac{2\pi}{3} \left(n+\frac{1}{2}\right)$,
separating regions of parameter space where the Polyakov loop angle
$\langle\varphi\rangle$ falls in different $Z(3)$ sectors. 
The resulting phase diagram in the
$(\mu_I,T)$ plane is periodic  and symmetric about $\bmu_I^c$, 
as depicted in Fig.~1 for $N_f=3$.
The $Z(3)$ transitions are first order for high temperatures, terminating
at some critical temperature $T_c$ which coincides with 
the deconfinement line. This endpoint also belongs to the 
3d Ising universality class \cite{fp}. The curves in Fig.~1 correspond to the deconfinement
line continued to imaginary chemical potential, $T_c(\mu_I)$. 
We will map out these curves as well as the
location of the critical endpoint
as a function of quark mass. Their relation to real $\mu$ is 
discussed in detail in the next section.
\begin{figure}[tb]
\centerline{\epsfxsize=8cm\epsfbox{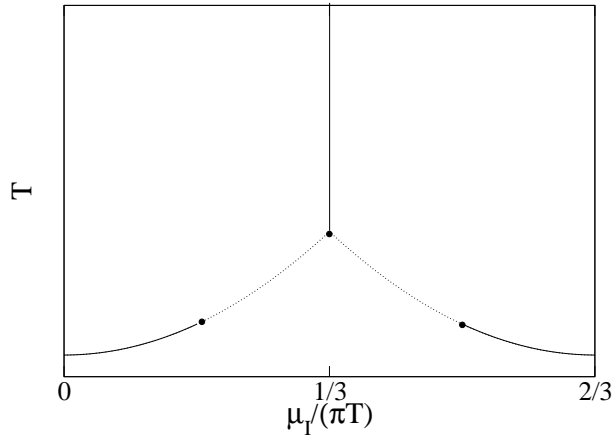}}
\caption[]{\em Schematic phase diagram in the $(\mu_I,T)$ plane. Solid lines
mark first order phase transitions, dotted ones crossover.
The vertical line corresponds to a $Z(3)$ transition and 
the curves to the deconfinement/chiral transition at imaginary $\mu$.
Both terminate in critical points, each belonging to the 3d Ising universality class.}
\label{schem}
\end{figure}

On the lattice, phase transitions can be studied
by measuring fluctuations of gauge invariant operators,
\be
\delta \op\equiv \op-\langle \op \rangle,
\qquad \op=\frac{1}{VN_t}\sum_{\bfx,t}\op(x),
\ee
where we use the plaquette, the chiral condensate and the modulus of the Polyakov
loop for $\op(x)$. A transition region is signalled by a peak in the susceptibilities
\be
\chi=V N_t \langle (\delta\op)^2 \rangle,
\ee
whose maximum implicitly defines the critical parameters, 
$\chi_{max}=\chi(\mu_c,\beta_c)$.
In a finite volume, the susceptibility is always an analytic 
function of the parameters
of the theory, even in the presence of a phase transition. The latter
reveals itself by a divergence of $\chi_{max}$ in the infinite volume limit,
whereas $\chi_{max}$ stays finite in the case of a crossover.
This fact was used in \cite{fp} to show that, on any large but finite volume,
the pseudo-critical coupling can be represented by a Taylor series in $\mu^2$.
Here we also wish to study the quark mass dependence, and thus consider an 
additional expansion
in $m$ about the $\mu=0$ chiral critical point $m_c(0)$. 
Hence we have for the pseudo-critical coupling on a finite volume
\be
\beta_c(a\mu,am)=\sum_{k,l=0} c_{kl}\, (a\mu)^{2k}\, (am-am_c(0))^l.
\label{beta}
\ee
With the help of the two-loop lattice beta-function, the critical coupling
can be converted to the critical temperature $T_c(m,\mu)$.
In our previous work \cite{fp} we demonstrated by simulations that 
the $\mu$-series converges fast and the critical coupling viz. temperature 
are well 
described by the leading $\mu^2$-term. 
Analytic continuation between real and 
imaginary chemical potential is then trivial. 

Following \cite{lom,hlp2,fp}, 
our strategy thus consists of measuring observables at $(\bmu_R=0,\bmu_I\neq 0)$,
fitting them by a
Taylor series in $\bmu_I^2$ and 
then continuing the truncated Taylor series to real $\bmu$.
The $Z(3)$-transition closest to the origin, at $\bmu_I^c=\frac{\pi}{3}$,
defines the convergence radius of the expansion and 
limits the prospects of analytic continuation.
In physical units this corresponds to $\mu_B \lsim 500$~MeV.

\section{\label{gen} Qualitative features of the phase diagram}

In principle, as proposed in \cite{fp}, the order of the phase transition, 
and hence the location of
the critical point, can be determined from a finite size scaling analysis
of the critical coupling $\beta_c(V)$ itself,  
which attains its infinite volume limit as 
\be \label{besc}
(\beta_c(V,\bmu)-\beta_c(\infty,\bmu))\sim \mbox{const}~V^{-\sigma(\bmu)},
\ee  
where $\sigma=1$ for a first order phase transition, $\sigma=1/d\nu < 1$
for a second order phase transition, and $\sigma=0$ for a crossover.
The critical endpoint on the curve is then defined by $\beta^*=\beta_c(\bmu^*)$,
for which the transition is of second order.
However, such an analysis is not practical for analytic continuation.
In a Taylor expansion in $\bmu$, the volume dependence resides in the 
coefficients,
\be
\beta_c(\bmu,V)=\beta_c(0,V)+ c_1(V)\bmu^2 + c_2(V)\bmu^4+\ldots
\ee
Clearly, a $\bmu$-dependent finite volume behavior as in \eq (\ref{besc})
cannot in general be well approximated by only a few terms of this series.

\begin{figure}[tb]
\leavevmode
\epsfysize=5.5cm\epsfbox{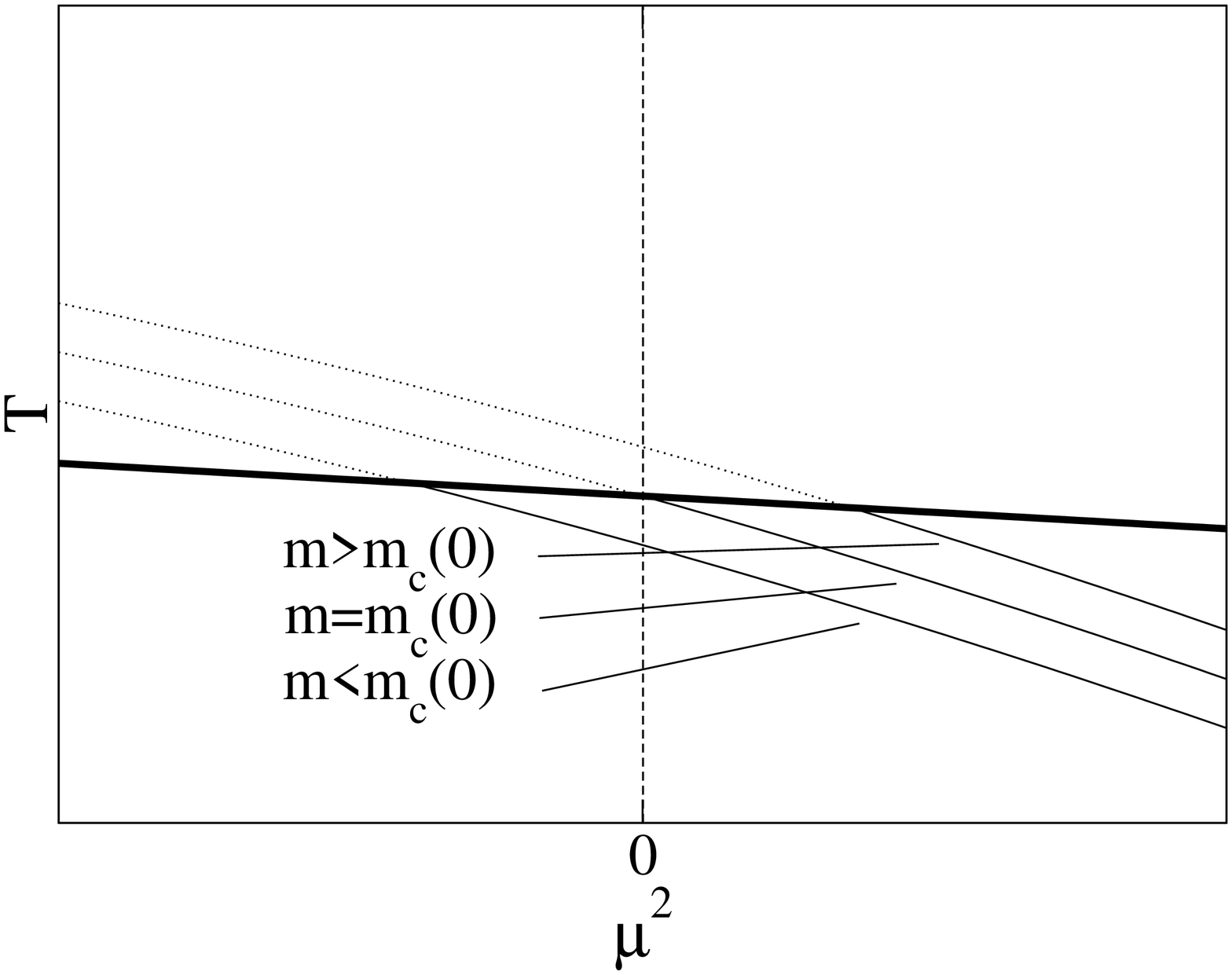}
\hspace*{1.0cm}
\epsfysize=5.5cm\epsfbox{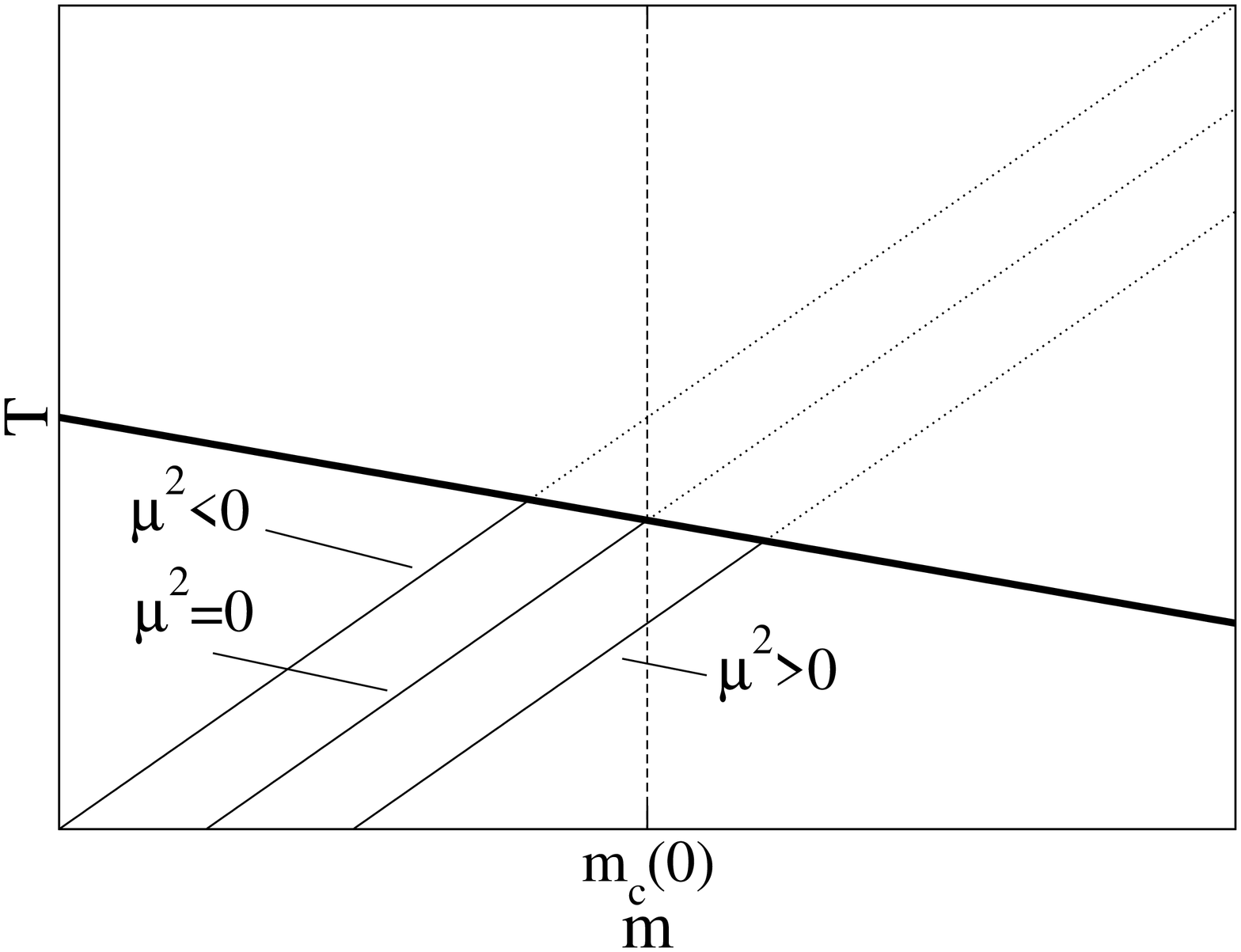}
\caption[]{Left: {\em critical lines in the $(T,\mu^2)$-plane for different quark masses $m$.
The bold curve $T^*(\mu)$ characterizes 
second-order transitions, separating the crossover and the first order regimes.}
Right: {\em critical lines in the $(T,m)$-plane for different chemical potentials
$\mu^2$.
The bold curve represents $T^*(m)$.}}
\label{eline}
\end{figure}

\begin{figure}
\vspace*{1cm}
\begin{center}
\leavevmode
\epsfysize=5.5cm
\epsffile{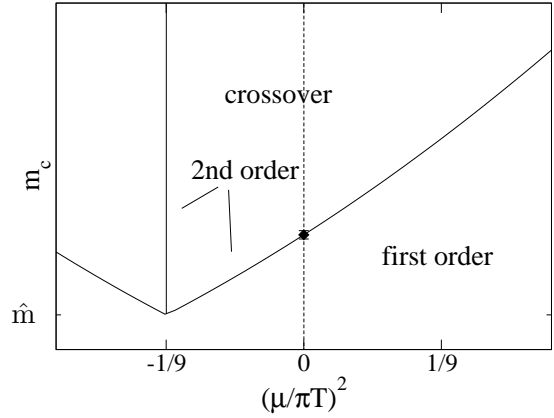}
\put(-205,35){\small $\rm{\hat{m}}$}
\end{center}
\caption[]{\em Schematic line of critical quark mass separating the 
first order and crossover region.
The line is constrained by the $\mu=0$ 
data point (diamond, \cite{kls}) and the fact that for 
$m=0$ the phase transition has to be first order
for all imaginary $\mu^2<0$, implying that intersection with the $Z(3)$-line 
happens at some quark mass $\hat{m}\geq 0$.} 
\label{mc_schem}
\end{figure}

Fortunately, the problem can be approached differently by considering 
variations of the quark mass, as outlined in \cite{fp2}. 
In this case we have a three-dimensional 
parameter space $\{T,\bmu, m\}$. The critical temperature
$T_c(\mu,m)$ now describes a surface in this space, and 
the critical endpoint traces out a line $T^*(m)=T_c(\mu^*(m),m)$, or equivalently
$T^*(\mu)=T_c(m_c(\mu),\mu)$, on this surface. 
Projections of this situation onto the $(T,\mu),(T,m)$ and $(m,\mu)$-planes 
are shown schematically in \fig \ref{eline} and \ref{mc_schem}. 
The bottom line in \fig \ref{eline} (left) corresponds to the situation
depicted in \fig \ref{schem}, for some quark mass $m<m_c(0)$.
With increasing quark mass, the critical endpoint of the deconfinement line
in \fig \ref{schem} moves towards $\mu=0$, which it hits for $m=m_c(0)$.
On the other hand, for decreasing quark mass
it moves to larger $\mu_I$, until it meets the $Z(3)$ endpoint at some mass
$\hat{m}$. For $m\leq\hat{m}$ the deconfinement and $Z(3)$ transition lines are
connected. 

This feature appears in \fig \ref{mc_schem} as the intersection of
$m_c(\mu)$ with the vertical $Z(3)$-line, which 
can be shown as follows.
In the chiral limit, the chiral condensate represents a true order 
parameter which is strictly zero in the deconfined phase, and there must be
a true phase transition for all values $\mu^2<0$.
Thus the line separating first order from crossover, $m_c(\mu)$, cannot hit
the negative $\bmu^2$-axis unless it has an unexpected non-analyticity there, 
implying the existence of some $\hat{m}\geq 0$.

Since the (pseudo-) critical line is analytic, so 
is the line of endpoints $T^*(\mu)$, and by elimination of $T$ 
the same holds for $m_c(\mu)$.
These are again smooth functions
with analytic continuations to imaginary $\mu$, which 
one may hope to describe well
in terms of only a few coefficients. In our practical calculation
we attempt to map out the phase diagram \fig \ref{mc_schem} by computing
the coefficients of
\be
a m_c(\bmu^2)=\sum_{n} c_n (a\mu)^{2n}\;.
\ee
Preliminary results for the leading coefficient as determined from
Taylor expanded reweighting, have been reported in \cite{cs}, and we will discuss this
result in comparison to ours in Sec.~\ref{dis}.

\subsection{\label{exp} Theoretical expectations}

Before continuing to describe our calculational tools, let us make a few 
remarks about what one would expect theoretically.
The analytic continuation approach has by now been tested for screening
masses in the plasma phase \cite{hlp2} as well as for $T_c(\mu)$ \cite{fp,el}. 
In \cite{hlp2} it was remarked that the screening masses are most ``naturally''
expanded in $(\mu/(\pi T))$, where ``natural'' means that the coefficients 
in such a series are of order one. 
The same observation is made regarding the critical temperature for the 
two flavor case. The result quoted in \cite{fp} can be rewritten as
\be
\frac{T_c(\mu)}{T_c(\mu=0)}= 1 - 0.500(67) \left(\frac{\mu}{\pi T_c}\right)^2,
\ee
where the coefficient is of order one. 
In thermal perturbation theory this is easy to understand, as in 
the imaginary time formalism one expands in terms of Matsubara modes and the chemical
potential always appears in this combination \cite{hlp2,vuo}. 
It is also transparent non-perturbatively 
in the case of an imaginary chemical potential
$\mu_I$:
the chemical potential gives an extra factor $\exp(\ii \mu_I/T)$ 
for the boundary
condition on the fermionic fields, so that it is equivalent to shifting
the Matsubara frequencies $(2 k + 1) \pi T$ by $\mu_I$. Hence the relevant
expansion parameter is the relative shift $(\mu_I/(\pi T))$.
We will empirically confirm this for the three flavor case, 
where we also measure the next-to-leading coefficient.

The same considerations apply to the quark mass expansion and provide
a reason for the near independence of $T_c(\mu)$ upon the light quark masses \cite{hk}.
Fermionic modes contribute with non-zero Matsubara frequencies,
and light quark masses are always negligible compared to
those modes, $(m/\pi T)\ll 1$. This is still the case for the strange quark
mass, and so we expect the curve $T_c(\mu)$ to be approximately  
the same for $N_f=3$ and $N_f=2+1$.  

For the critical quark mass one then similarly expects to have
\be \label{mcseries}
\frac{m_c(\mu)}{m_c(\mu=0)}=1 + c_1 \left(\frac{\mu}{\pi T_c}\right)^2+\ldots\;,
\ee
with $c_1$ of order one. The bare quark mass is not a physical quantity, but
depends on the lattice action. For example,
comparing calculations with p4-improved and unimproved staggered fermion actions,
one finds $m_c(0)|_{impr}\approx 0.25 m_c(0)|_{unimpr}$ \cite{kls}.
However, the mass renormalization should not be affected by $\mu\neq 0$, which
is just an external thermodynamic parameter without ultraviolet renormalization.
Multiplicative mass renormalization should therefore
cancel out in the ratio \eq (\ref{mcseries}), which,
up to additive corrections $\op(a^2)$, is
directly comparable between different lattice actions.

A remarkable finding for the critical temperature is that it is quite
accurately described by the leading $\mu^2$ term, at least up to 
$|\bmu|=\bmu_I^c$, where for imaginary $\mu$ the $Z(3)$ transition occurs.
The same was found for screening masses \cite{hlp2} and recently also 
for the pressure \cite{alt1}. We thus expect similar
behavior for $m_c(\mu)$, which will be confirmed by our 
simulations. 
If $m_c(\mu)$ is well described by the leading term,
its intersection with the $Z(3)$-line at a quark mass value $\hat{m}\geq 0$, 
as described in the previous section,
furthermore implies an upper bound for $c_1$. 
From Eq.(\ref{mcseries}) one gets $(1 - c_1 \left(\frac{\mu_I^c}{\pi T}\right)^2) \geq 0$,
or $c_1 \leq 9$.

\section{\label{meth} Cumulant ratio and finite volume scaling}

In order to find the boundary between the first order and crossover regime along 
the critical line,
we use the Binder cumulant \cite{bin} of the chiral condensate,
\be 
B_4=\frac{\langle(\delta\bar{\psi}\psi)^4\rangle}
{\langle(\delta\bar{\psi}\psi)^2\rangle^2}.
\ee
In the infinite volume limit this quantity assumes a universal value
at a critical point. 
In particular, this observable was used in \cite{kls} to locate the chiral critical point
at $\mu=0$, $a m_c(0)=0.0331(12)$ for staggered fermions on an $N_t=4$ lattice, and to identify
its universality class as that of the 3d Ising model, for
which $B_4\approx 1.604$. 
On a finite volume, this value receives corrections. It also receives corrections
away from the critical point, which are positive for crossover and negative for
first order behavior. 
Cumulants calculated on increasing lattice sizes for different parameters
will intersect at some pseudo-critical value of the parameters, 
with the $B_4$-value at the
intersection point converging towards its universal value.

In order to explicitly assess the quark mass and $\mu$-dependence, 
we fit our data by a Taylor expansion
about the $\mu=0$ critical point,
\be \label{Bexp}
B_4(m,\mu)=\sum_{n,l} b_{nl}\, (am-am_c(0))^n(a\mu)^{2l}\;,
\ee
with $b_{00}(V\rightarrow \infty)=1.604$.
This observable can also be directly related to the critical line in the phase diagram
\fig \ref{mc_schem}. At the expansion point $m_c(0)$ we have $B_4=b_{00}$, and 
this value is maintained along the line $m_c(\mu)$, which is a line of constant $B_4$.
This line is implicitly defined by the equation
\be
B_4(m_c(\mu),\mu)=b_{00},
\ee 
and in particular one obtains the coefficient $c_1$ of \eq (\ref{mcseries})
through the chain rule
\be \label{der}
\frac{dam_c}{d(a\mu)^2}=-\frac{\partial B_4}{\partial (a\mu)^2}
\left(\frac{\partial B_4}{\partial am}\right)^{-1}=-\frac{b_{01}}{b_{10}}\;.
\ee

If the volumes are large enough, the approach to the thermodynamic 
limit is governed by universality.
In this case the volume dependence hidden in the 
coefficients of the series can be made explicit. Approaching the critical
endpoint, the correlation length diverges
as $\xi\sim r^{-\nu}$, where $r$ is the distance to the critical point
in the plane of temperature and magnetic field-like variables, and $\nu=0.63$ 
for the Ising universality class. In practice, we first find
$\beta_c$ for a given pair $(m,\mu)$, 
and then compute $B_4$ for those values of the couplings.
Since $\beta=\beta_c$ always, we thus have $r=|m-m_c(\mu)|$. 
$B_4$ is a function of the dimensionless ratio $L/\xi$, or equivalently 
$(L/\xi)^{1/\nu}$. Hence  one expects
the scaling behavior
\be \label{scale}
B_4\left((L/\xi)^{1/\nu}\right)=B_4\left(L^{1/\nu}(am-am_c(\mu)\right)\;.
\ee

\section{\label{num} Numerical results}

In our simulations we consider QCD with the Wilson gauge action and three 
degenerate flavors
of staggered fermions, with
bare quark masses in the range $0.025 < am < 0.04$. 
We monitor finite volume scaling behavior using
three lattice sizes, $8^3\times 4$, $10^3\times 4$ and $12^3\times 4$.
The Monte Carlo employs the R-algorithm~\cite{R-alg} 
with a step size $\delta\tau = 0.02$, which is
sufficiently small for the systematic errors ${\cal O}(\delta\tau^2)$ to be
negligible compared to our statistical errors. 
For each simulated parameter set we accumulate $10k-70k$ 
unit-length trajectories, measuring
the gauge action and the Polyakov loop and estimating the first four 
powers of the chiral condensate after each 
trajectory.
The pseudo-critical values $\beta_c(a\mu_I)$ are obtained 
from a range of typically
$4$ simulated $\beta$-values by means of the Ferrenberg-Swendsen reweighting 
method~\cite{FS}. Hence, every data point in the following figures for 
the critical coupling and the cumulant ratio typically consists of over
$100k$ trajectories.

\subsection{The critical line $T_c(\mu)$}

The calculation of the critical line proceeds as in the 
two flavor case \cite{fp}. The critical coupling $\beta_c$ was determined by
finding a peak in the plaquette susceptibility, and we have checked 
that the chiral condensate and the Polyakov loop give consistent 
results. Our first task then is to determine the coefficients 
in the Taylor expansion \eq (\ref{beta}).
 
In Table \ref{fits} we give an exhaustive list of all possible three, four and six
parameter fits to our data. For the expansion point in the quark mass, we have
chosen $am_c(0)=0.0323$, which will be the result obtained in Sec.~\ref{cum}.
Apart from resolving the leading linear quark mass and quadratic chemical 
potential dependence, our statistics is now also large enough to permit
some statements concerning the next-to-leading terms. On our $L=10$ lattice 
we studied the largest $a\mu_I$, and consequently get the most constrained
fits for the $\mu^4$-term. Note that on this volume a
quartic term is required to fit the data, while the other possibilities give
significantly worse fits.
The situation on the other volumes is consistent with this. The best four parameter fit is 
in all cases the one with a $\mu^4$-term. The other options give 
coefficients that are either consistent with zero within 1.5 standard deviations,
or inconsistent between the different volumes. On the other hand, six parameter fits
do not significantly reduce the $\chi^2$, and thus are not fully constrained yet.

\begin{figure}[t]
\vspace*{1cm}
\centerline{
\epsfxsize=7.5cm\hspace*{1cm}\epsfbox{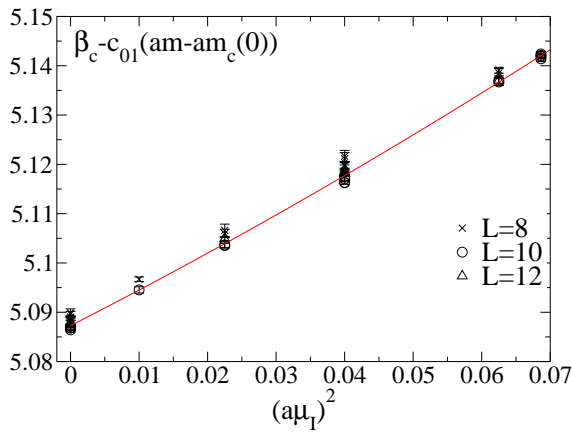}}
\caption[a]{
\em Combined
data for $L=8-12$ and various quark masses. Data for different $am$ 
are shifted to $a m_c(0)$ according to the 
best fit of Table 1, which is also shown. A weak $\mu^4$-dependence is visible.}
\label{tc}
\end{figure}

Comparing the coefficients of the fits including the quartic term between the
volumes, we observe that the present statistics is unable to resolve systematic
finite volume effects, all volumes being compatible within one standard deviation.
It is then expedient to further constrain the fit parameters by fitting all
volumes together. We use the best four parameter fit highlighted in the table as our final
result for the critical coupling, which is shown in \fig \ref{tc} as a function of 
$\mu^2$.

\begin{table}[t]
\hspace*{-1cm}
\begin{tabular}{|c|*{7}{r@{.}l|}l|}
\hline
\hline
$L$ &
\multicolumn{2}{c|}{$c_{00}=\beta_c(0,0)$} &
\multicolumn{2}{c|}{$c_{10}, (\mu^2)$} &
\multicolumn{2}{c|}{$c_{20}, (\mu^4)$} &
\multicolumn{2}{c|}{$c_{01}, (m)$} &
\multicolumn{2}{c|}{$c_{02}, (m^2)$} &
\multicolumn{2}{c|}{$c_{11}, (\mu^2 m)$} &
\multicolumn{2}{c|}{$\chi^2/{\rm dof}$} \\
\hline
 8 & 5&1450(2) & 0&790(8)  & \none    & 1&748(26) & \none    & \none        & 0&94 \\
 8 & 5&1451(2) & 0&739(26) & 0&93(46) & 1&735(27) & \none    & \none        & 0&66 \\
 8 & 5&1449(3) & 0&791(8)  & \none    & 1&745(27) & 2&3(6.3) & \none        & 1&01 \\
 8 & 5&1450(2) & 0&789(8)  & \none    & 1&751(31) & \none    & -0&085(0.40) & 1&02 \\
 8 & 5&1449(4) & 0&740(27) & 0&99(48) & 1&712(44) & 6&8(9.9) & 0&38(0.63)   & 0&75 \\
10 & 5&1449(2) & 0&800(5)  & \none    & 1&831(23) & \none    & \none        & 7&66 \\ 
10 & 5&1457(2) & 0&667(15) & 1&98(22) & 1&818(23) & \none    & \none        & 1&21 \\
10 & 5&1457(2) & 0&798(5)  & \none    & 1&808(23) &-28&2(4.6)& \none        & 5&41 \\
10 & 5&1449(2) & 0&798(5)  & \none    & 1&883(40) & \none    & -0&38(25)    & 8&04 \\
10 & 5&1458(2) & 0&679(17) & 1&78(25) & 1&820(42) & -8&6(5.6)& -0&07(26)    & 1&21 \\
12 & 5&1455(4) & 0&764(13) & \none    & 1&770(44) & \none    & \none        & 1&04 \\
12 & 5&1457(5) & 0&721(31) & 0&94(63) & 1&788(46) & \none    & \none        & 0&44 \\
12 & 5&1454(5) & 0&763(13) & \none    & 1&784(57) & 7&5(18.9)& \none        & 1&48 \\
12 & 5&1443(8) & 0&791(21) & \none    & 2&28(31)  & \none    & -2&5(1.5)    & 0&12 \\
8-12 & 5&1442(1) & 0&7943(33) & \none & 1&796(16) & \none & \none & 4&10 \\
\hline
8-12 & 5&1453(1) & 0&705(10)  & 1&46(15) & 1&780(16) & \none & \none & 1&43 \\
\hline
8-12 & 5&1453(2) & 0&7917(34) & \none & 1&792(16) & -14&2(3.3) & \none & 3&68 \\
8-12 & 5&1448(1) & 0&7940(34) & \none & 1&807(23) & \none & -0&1(16) & 4&20 \\
8-12 & 5&1457(2) & 0&705(10)  & 1&43(16) & 1&767(26) & -10&1(3.9) & 0&10(19) & 1&17 \\
\hline
\hline
\end{tabular}
\caption{ \label{fits} 
  {\em 
Fits of the Taylor expansion $\beta_c(m,\mu)$,
\eq (\ref{beta}), to our data.}}
\end{table}

We conclude that we
have a signal for a $\mu^4$ contribution to the critical coupling.
This is a result of having more accurate data and does not invalidate
our earlier observation that the line is well described by the leading term. 
E.g.~at $\mu_I^c$ the contribution of this term to the critical coupling is
only $\sim 0.1\%$. Converting to continuum units by means of the two-loop
beta-function as in \cite{fp}, we thus obtain for the critical line in three flavor QCD
\be \label{tc1}
\frac{T_c(\mu,m)}{T_c(0,m_c(0))}= 1 
+ 1.937(17) \left(\frac{m-m_c(0)}{\pi T_c}\right) 
-0.602(9)\left(\frac{\mu}{\pi T_c}\right)^2
+0.23(9)\left(\frac{\mu}{\pi T_c}\right)^4.
\ee
$T_c$ on the right side of this and the following two equations is meant to be the same
as in the denominator on the left, which plays a role
for the coefficient of next-to-leading terms.
Note that the mixing of the errors on the parameters in the critical coupling through
standard error propagation drowns out the $\mu^4$-signal in continuum units.
Since we do not yet have a signal for a mixed $(m\mu^2)$-term, the quark mass and chemical
potential dependence are separately consistent with
\ba
\frac{T_c(\mu,m)}{T_c(\mu,m_c(0))}&=& 1
+ 0.617(5) \left(\frac{m-m_c(0)}{T_c}\right)\;,\nn\\
\frac{T_c(\mu,m)}{T_c(0,m)}&=& 1
- 0.00678(10) \left(\frac{\mu_B}{T_c}\right)^2
+0.000029(11)\left(\frac{\mu_B}{T_c}\right)^4.
\label{tc2}
\ea
A mixed dependence only appears in higher orders, having no effect at our present accuracy.
This explains why critical lines obtained previously for various different quark masses
agree so well \cite{lap}.

\begin{figure}[t]
\vspace*{1cm}
\leavevmode
\epsfysize=5.5cm\epsfbox{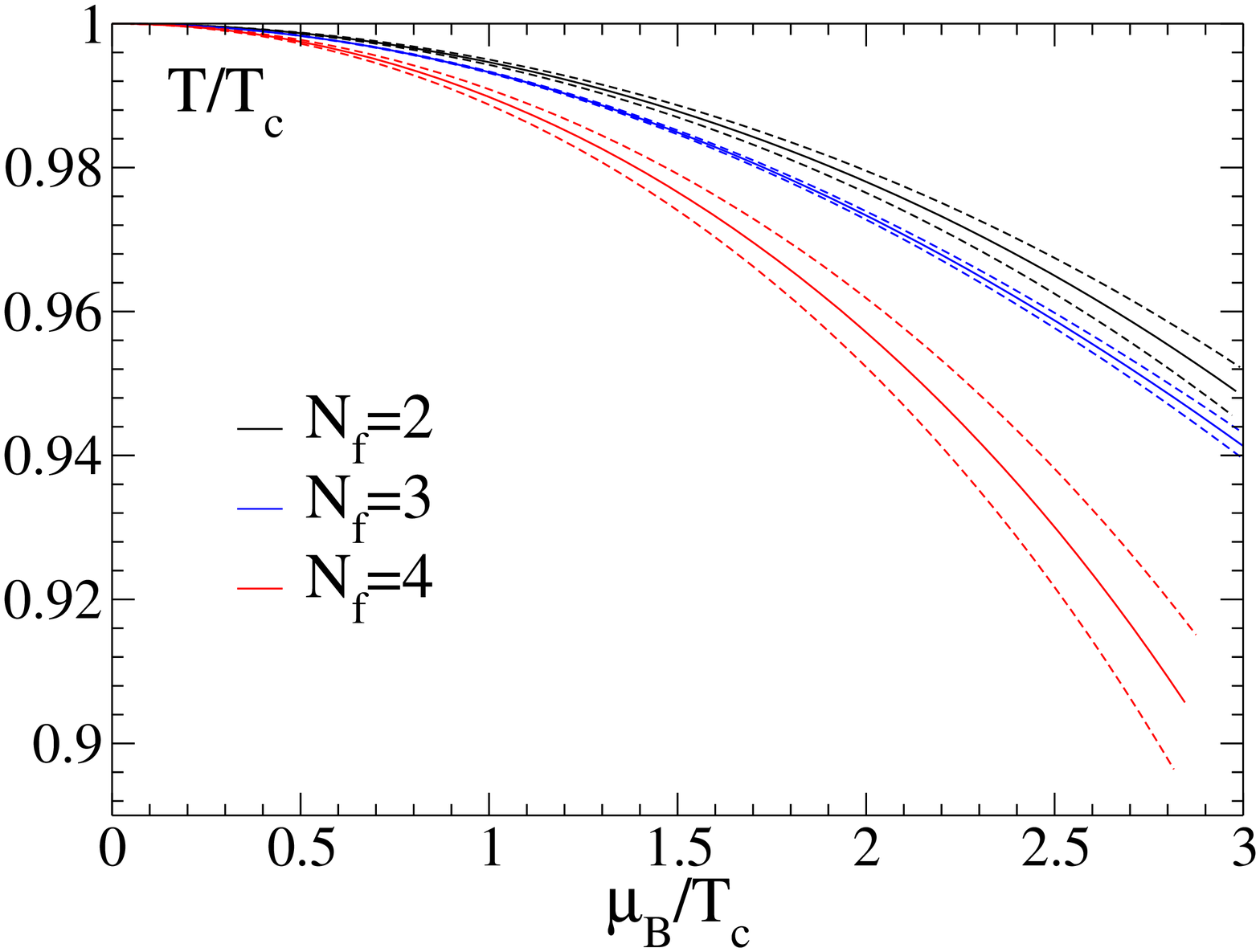}
\hspace*{1cm}
\epsfysize=5.5cm\epsfbox{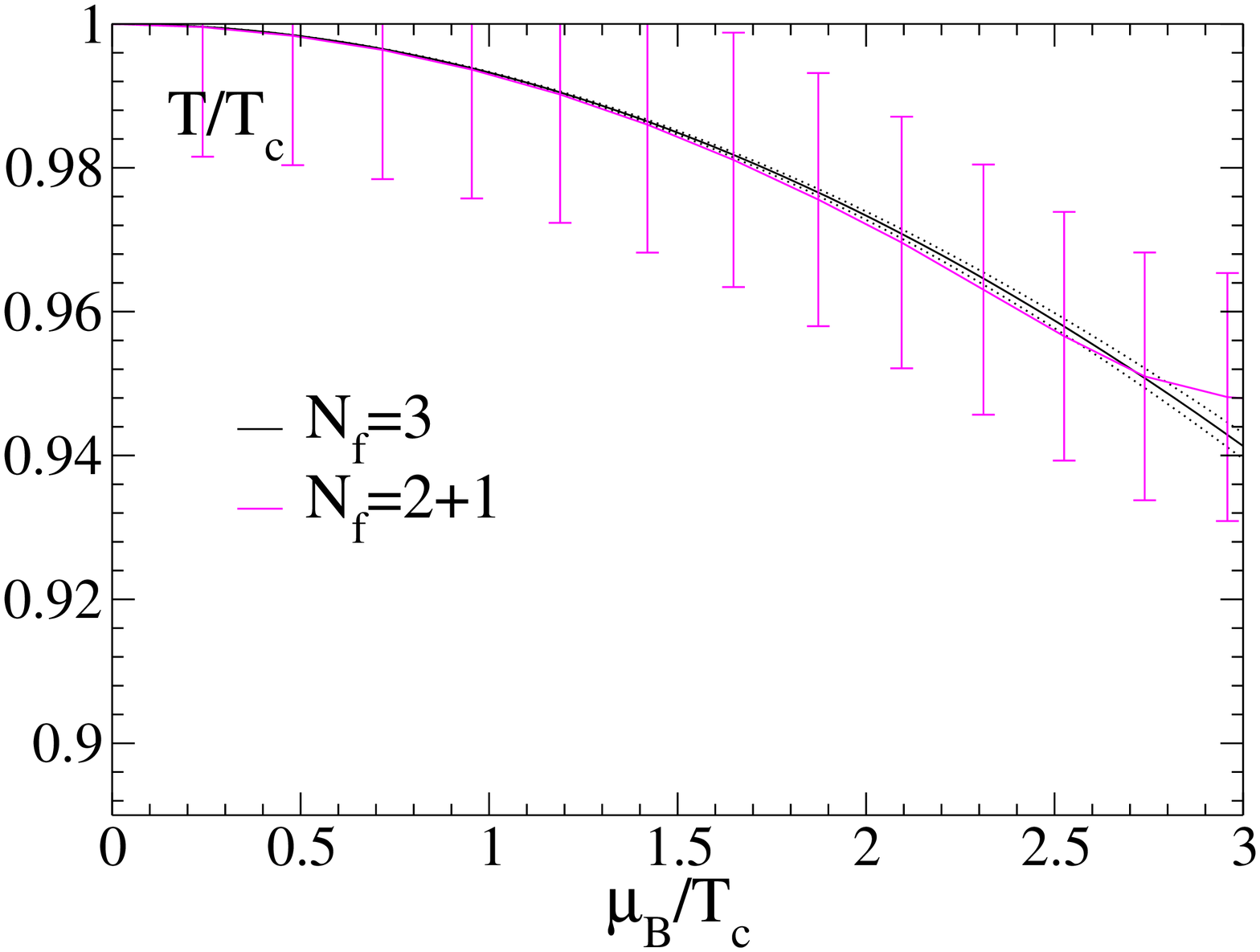}
\caption[a]{
Left: {\em One sigma error bands on $T_c(\mu_B)$ 
for different $N_f$ ($N_f=4$ from \cite{el}). Only the $N_f=3$ calculation is 
accurate enough to include a quartic term.} Right: {\em Comparison of $N_f=3$ with $N_f=2+1$
from \cite{fk2}.}
}
\label{tc_nf}
\end{figure}

We may then directly compare our result with existing ones for 
$N_f=2$ \cite{fp} and $N_f=4$ \cite{el} in \fig \ref{tc_nf}.
As one would expect, our result falls between these two. Note, however,
that these earlier results were not sensitive to a $\mu^4$-term, which 
makes itself felt at the right end of the interval and also lowers the $\mu^2$-coefficient,
cf.~Table \ref{fits}.
Also shown in the figure is the result for $N_f=2+1$ as obtained by 
reweighting \cite{fk2}. In accordance with our expectations from 
Sec.~\ref{exp}, due to the quark mass independence of the critical line
this result is practically identical to the one for $N_f=3$.

\subsection{\label{probs} First order vs. crossover and error estimates}

Before presenting our results for the cumulant ratio, we make some remarks concerning 
the considerable technical difficulty of these measurements.
Inspection of the Monte Carlo history of an observable
over a sufficiently long Monte Carlo time reveals that the tunneling frequency between
the different vacua is very low: observing only one crossing 
per a few thousand trajectories is typical. This
is expected in a first order regime, where tunneling is suppressed by a potential barrier,
but the same observation is made in the crossover regime. 

\begin{figure}[t]
\leavevmode
\vspace*{1cm}
\epsfysize=5.5cm\epsfbox{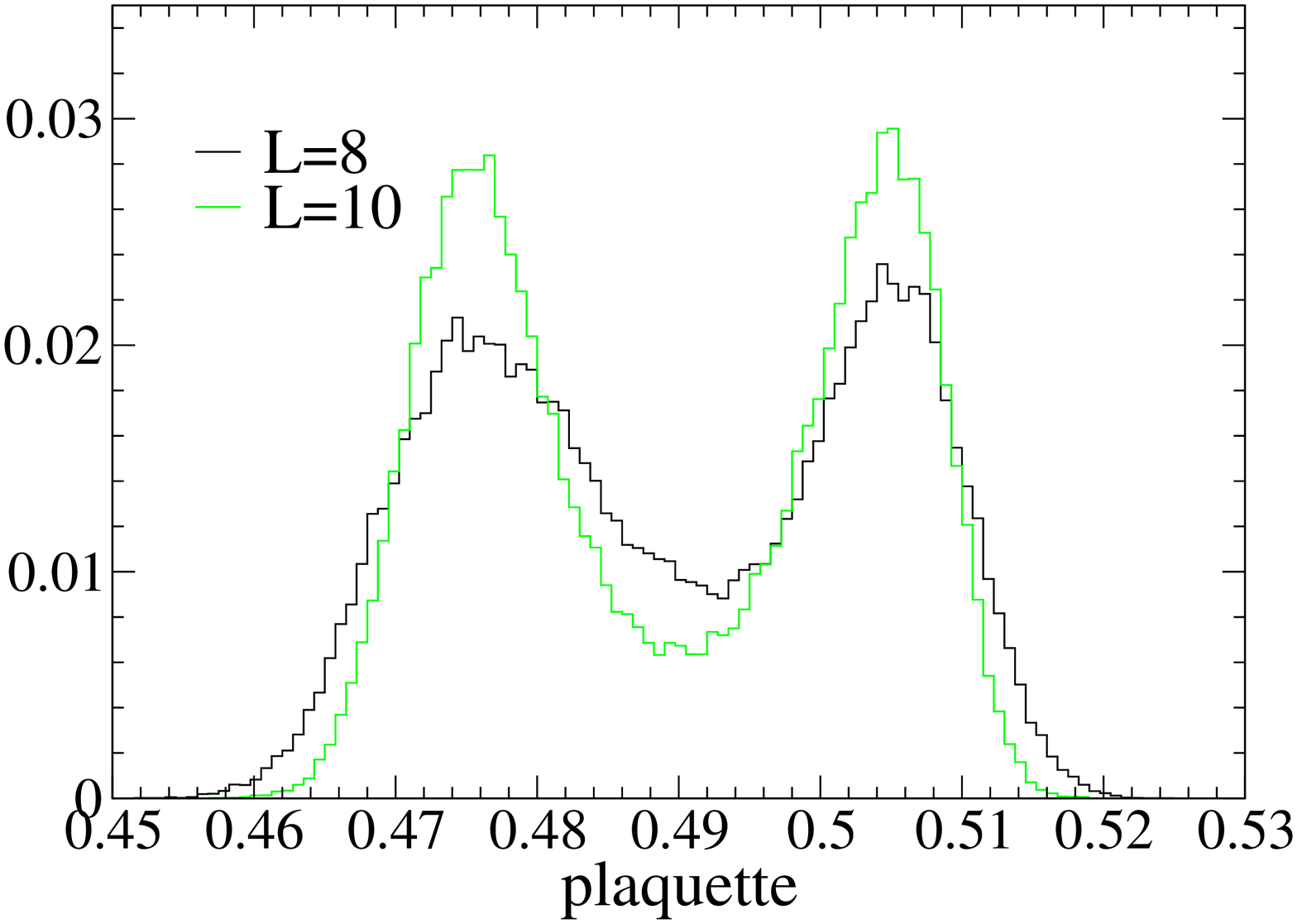}
\hspace*{1cm}
\epsfysize=5.5cm\epsfbox{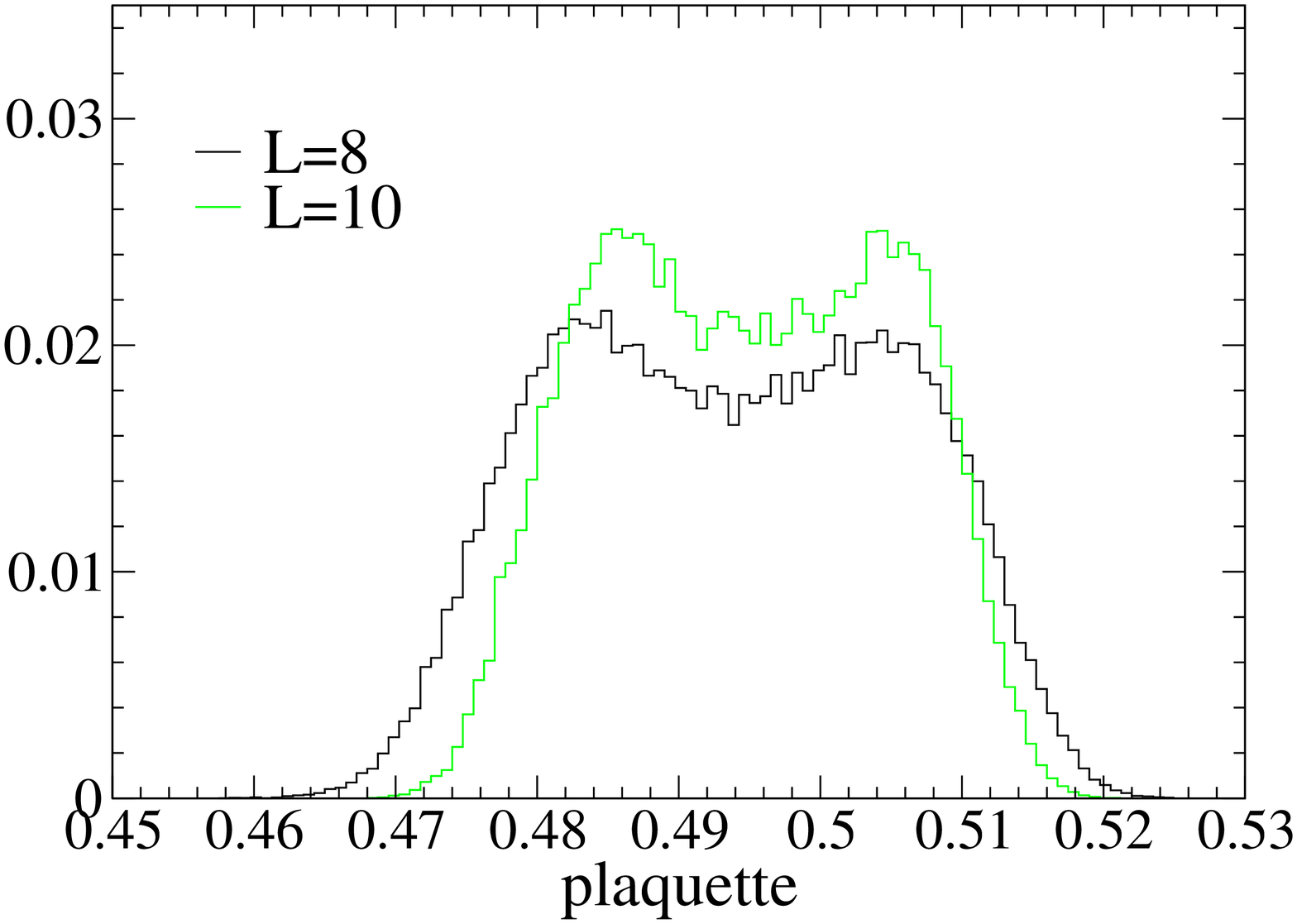}
\caption[a]
{\em Distribution of plaquette values at $\beta_c$ for $L=8,10$ and $\mu=0$.
{\rm Left:} First order transition, $am=0.025$.
{\rm Right:} Crossover, $am=0.04$.
}
\label{hist}
\end{figure}

The reason for this behavior is the fact that, on the lattice sizes used here, the 
probability distributions for measurements at the critical coupling
$\beta_c(m,\mu)$ have not yet reached their asymptotic scaling regime. 
This is illustrated in
\fig \ref{hist}, where we show the distributions of plaquette values on two volumes for 
a point each in the first order and crossover regimes.
In accordance with expectation, the first order region displays a two peak structure and 
tunneling gets more suppressed on a larger volume. 
In the crossover region we observe accordingly
a merging of the two peak structure with increasing volume. However, this merging to the 
asymptotic Gaussian distribution is not yet complete, and a remnant of the two-peak structure
can be clearly identified. The displayed parameter values are deep in the 
crossover region, and the
situation gets only worse closer to the critical point. 
This is a well known difficulty
in the investigation of phase transitions. 

On the other hand, the value of $B_4$ and its statistical error are driven by the number of tunnelings
rather than the total number of measurements. 
Essentially the observable distinguishes between crossover and
a first order transition by picking up the difference in the frequency of tunnelings.
This leads to a much slower reduction of error bars than in the case
of the critical couplings, where only the change of the observable 
between the two phases is needed, for which the
number of measurements is relevant. 
Hence, too short Monte Carlo runs with less than a few tens of tunnelings tend to 
underestimate the statistical error on $B_4$. 
More dangerous is the finite volume remnant of tunneling suppression
in the crossover regime which can, 
for too short runs and combined with too small an error estimate, 
lead to an underestimate of $B_4$ and hence to 
misidentifying a crossover as a weakly first order signal.
  
In light of this, we can only be fully confident of our $B_4$ error estimates
for $L=8$ lattices, where tunneling is faster and we have the longest Monte Carlo runs.
On this volume we obtain a significant result for the $\mu$-dependence,
whereas for $L=10,12$ the signal is hidden in the noise. These
volumes will be mainly used for consistency and scaling checks.

\subsection{\label{cum} The cumulant ratio as function of $m$ and $\mu$}

\begin{figure}[tb]
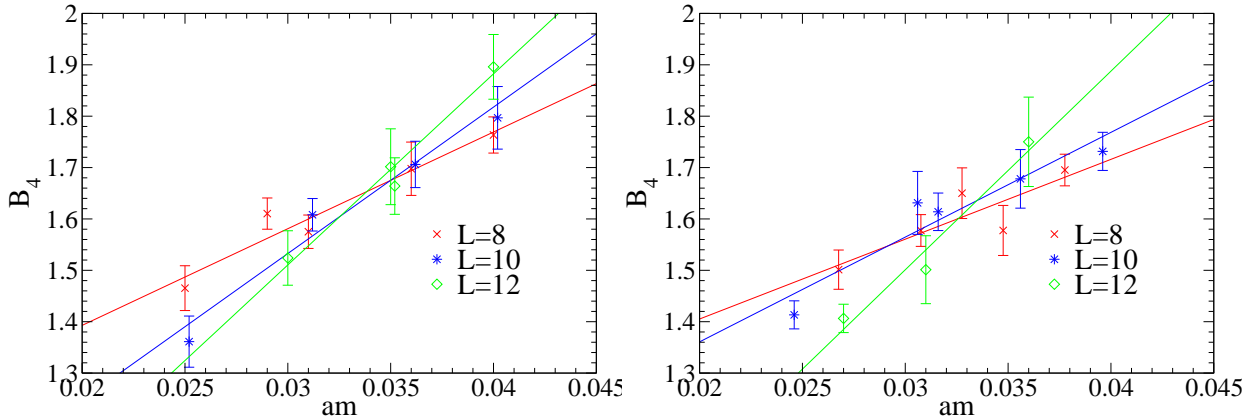

\leavevmode
\epsfysize=5.5cm\epsfbox{B4_mu0.0eps}
\epsfysize=5.5cm\epsfbox{B4_mu0.2eps}
\caption[]{\em Binder cumulant for $\beta=\beta_c$ as function of quark mass
for $a\mu_I=0$ (left) and $a\mu_I=0.2$ (right).}
\label{B4_v}
\end{figure}

Following Sec.~\ref{meth}, we proceed to discuss our measurements of the cumulant
ratio $B_4$ along the critical line in order to determine its endpoint and its
quark mass dependence. 
Our current accuracy constrains only 
the leading terms $\op(am,(a\mu)^2)$ in the Taylor expansion of $B_4$.
To begin, we perform an analysis analogous to the one in \cite{kls}. For a fixed value
of $a\mu_I$, we measure $B_4$ along the critical line $\beta_c(am,a\mu_I)$,
cf.~\fig \ref{eline} (right). The critical quark mass separating first
order from crossover behavior is then extracted
from the intersection of $B_4$ measured
on different volumes. This is shown in \fig \ref{B4_v} for $a\mu_I=0$ and $a\mu_I=0.2$.
Our results for $a\mu_I=0$ are in full agreement 
with those reported in \cite{kls}, serving as a check of the analysis.
The volume dependence appears to be moderate, and for 
the intersection point between the larger volumes we get $am_c(0)\approx 0.033$,
compared to $am_c(0)=0.0331(12)$ \cite{kls}. However, practically 
the same result is obtained for $a\mu_I=0.2$, pointing to a very weak $\mu$-dependence
of $B_4$. Indeed, plotting our data for fixed $am$ as a 
function of $(a\mu_I)^2$, no structure beyond noise is apparent to the eye.

\begin{figure}[tb]
\begin{center}
\epsfysize=6.5cm\epsfbox{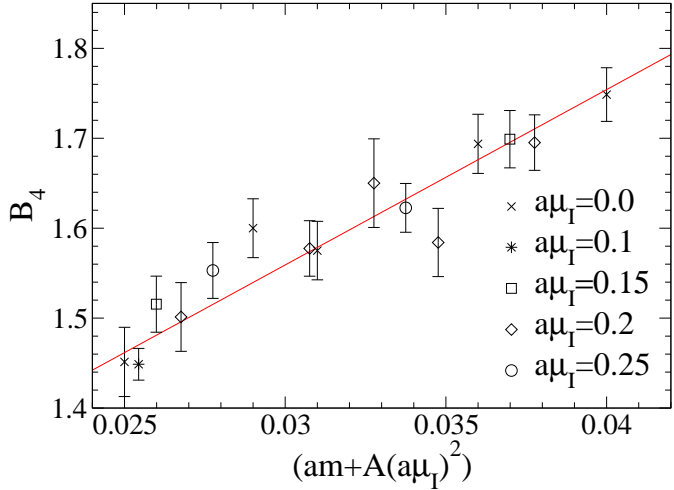}
\caption[]{{\em Binder cumulant in the $(am,a\mu_I)$-plane for $L=8$,
the line represents the fit given in Table \ref{B4-fit}.}}
\label{B4_8}
\end{center}
\end{figure}

In order to obtain better accuracy we modify our analysis.
Let us rewrite the leading terms of the Taylor expansion \eq (\ref{Bexp}) as
\be \label{b4}
B_4(am,a\mu)=1.604 + B\left(am-am_c(0) - A(a\mu)^2\right)\;,
\ee
where we have traded the parameters $\{b_{00},b_{01},b_{10}\}$ for 
$\{m_c(0),A,B\}$. With the constant fixed to its infinite volume value, finite
volume corrections to $b_{00}$ will now show up in $m_c(0)$, 
which can be compared with the previous result.
In this form we can collapse all our data obtained for various
pairings $(am,a\mu_I)$ into one plot and fit them by a single three parameter fit.
Finally, $d(am_c)/d(a\mu)^2$, as in \eq (\ref{der}),
is now immediately given by the fit parameter $A$.
\fig\ref{B4_8} shows all $L=8$ data combined in this way together with the best
fit. The fit results for all
volumes are displayed in Table \ref{B4-fit}.

\begin{table}[t]
\begin{center}
\begin{tabular}{|c|*{4}{r@{.}l|}l|}
\hline
\hline
$L$ &
\multicolumn{2}{c|}{$am_c(0)$} &
\multicolumn{2}{c|}{$A$} &
\multicolumn{2}{c|}{$B$} &
\multicolumn{2}{c|}{$\chi^2/{\rm dof}$} \\
\hline
 8 & 0&0313(5) & \none    & 18&5(1.8) & 1&33\\
 8 & 0&0323(6) & 0&044(19) & 18&9(1.6)) & 1&00\\
10 & 0&0325(3) & \none     & 25&9(1.3) & 0&54\\
10 & 0&0320(5) & -0&010(10) & 25&4(2.1) & 0&53\\
12 & 0&0326(2) & \none      & 35&8(1.7) & 0&19\\
12 & 0&0325(4) & -0&008(18) & 35&0(2.5) & 0&22\\
16 & 0&0331(3) & \none      & 57&0(6.3) & 0&19\\
\hline
$L$ &
\multicolumn{2}{c|}{$am_c(0)$} &
\multicolumn{2}{c|}{$A$} &
\multicolumn{2}{c|}{$B/L^{1/\nu}$} &
\multicolumn{2}{c|}{$\chi^2/{\rm dof}$} \\
\hline
8-12 & 0&0323(3)& 0&0008(8)  & 0&67(3)    & 0&84\\
\hline
\hline
\end{tabular}
\caption{ \label{B4-fit}
  {\em
Fits of the Taylor expansion $B_4$,
\eq (\ref{b4}), to the data. $L=16$ data for $\mu=0$ are taken from \cite{kls}.}}
\end{center}
\end{table}

However, even after combining
all data on one volume, the $\mu^2$-coefficient $A$
is still only weakly constrained.
The data on the larger lattices are consistent with a negligible $\mu$-dependence, 
as is apparent by the acceptable fits obtained without such a term. 
Only on the $8^3$ lattice, for which we have the best
statistics, do the fits prefer a positive value of this quantity.

\begin{figure}[tb]
\leavevmode
\centerline{\epsfxsize=8cm\epsfbox{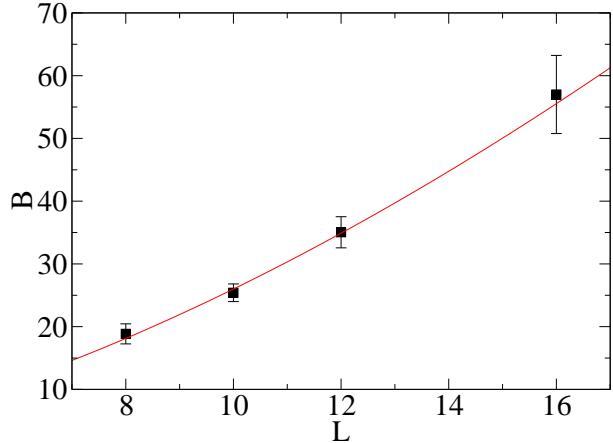}}
\caption[]{\em Finite volume scaling of the fit parameter $B$ from \eq (\ref{b4}).
The line represents a fit to $\sim L^{1/\nu}$, with $\nu=0.62(3)$, $\chi^2/{\rm dof}=0.2$.
}
\label{scale_plot}
\end{figure}

\begin{figure}[tb]
\vspace*{0.75cm}
\leavevmode
\centerline{\epsfxsize=8cm\epsfbox{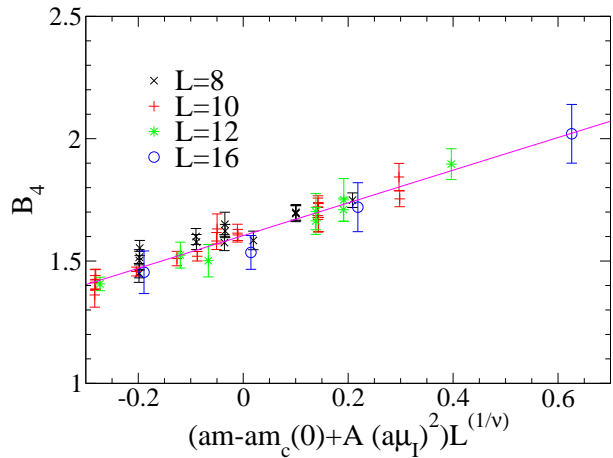}}
\caption[]{\em Combined fit for all volumes by fixing the scaling to the Ising form,
\eq (\ref{scale}).
}
\label{B4_all}
\end{figure}

Let us now try to combine the different volumes by
exploiting the fact that $m_c(0)$ appears close to its
infinite volume limit, and hence
$B_4$ should be close to the scaling region on the volumes considered.
In order to explicitly test for this, we plot the fit parameter $B$ against the volumes
for which it was obtained, and fit the data to the expected asymptotic
scaling behavior $B_4(L)\sim L^{1/\nu}$, cf.~\eq (\ref{scale}). 
For this purpose, we also use the $L=16$ data from \cite{kls}.
This is shown
in \fig \ref{scale_plot}, and the resulting $\nu=0.62(3)$ is indeed consistent with the 
Ising value $\nu \approx 0.63$.
Having thus established the explicit volume dependence of $B_4$ as in \eq (\ref{scale}), we
may combine all available volumes into one maximally constrained fit, in order to get higher 
accuracy. This is done in \fig \ref{B4_all}. The resulting parameter values
are given in the last line of Table \ref{B4-fit}.
Our result for the zero density critical quark masss 
then is $am_c(0)=0.0323(3)$, with a statistical error of 1\%, 
and in perfect agreement with the 
calculations at $\mu=0$ \cite{kls,nc}. 

\subsection{\label{mc} The line of critical endpoints}

The situation is less clear for the $\mu$-dependence, as Table \ref{B4-fit} shows.
The combined fit over all volumes does not constrain the parameter $A$ enough to 
yield a non-vanishing result. Clearly, this is due to the $L=10,12$ lattices,
whose results are consistent with zero, but whose negative central values neutralize
the significant answer obtained on $L=8$. Since these lattices only add noise to 
the determination of $A$, we thus quote the $L=8$ number as our tentative final result, 
\be \label{result1} 
\frac{m_c(\mu)}{m_c(\mu=0)}=1 + 0.84(36) \left(\frac{\mu}{\pi T}\right)^2,
\ee
with higher terms being smaller than our present error of 40\%. 
A check of the fit result is obtained by measuring $B_4$ for different quark masses along
the vertical $Z(3)$-line at $a\mu_I^c$, cf.~\fig \ref{mc_schem}, in order to determine
$\hat{m}$. We have done so on $L=10$ and find $0.029<a\hat{m}<0.032$, while \eq (\ref{result1})
in lattice units predicts $a\hat{m}=0.030$.

Note that, in terms of
our natural expansion units, the coefficient of interest is not unnaturally small.
Determining it to better accuracy is, however, a formidable numerical task that requires
computational resources on the largest scales available.  
A more conservative result is obtained by adding two standard deviations to the 
central value, resulting in a bound $c_1<1.6$ at 90\% confidence level. 

Taken at face value, \eq (\ref{result1}) tells us the critical 
bare quark mass for a given 
chemical potential as sketched in \fig \ref{mc_schem}, 
while its inverse yields the location of 
the critical endpoint for a given bare quark mass. 
The renormalization of the bare quark mass cancels in the ratio, so that
it should be independent of the lattice action chosen, up to additive 
cut-off effects.
Moreover, since 
in the mass range of interest the zero temperature pion mass $M_\pi^2\propto m$, we have
\be
(M^c_{\pi}(\mu))^2=(M_{\pi}^c(0))^2\frac{m_c(\mu)}{m_c(\mu=0)}\;,
\ee
where $M_{\pi}^c(0)\approx 290(190)$ MeV for unimproved 
(p4-improved (preliminary)) staggered
fermions, respectively \cite{kls,alt1}. 
These numbers highlight the strong need to eliminate cut-off effects
on $M_{\pi}^c(0)$. 

Another result involving only physical quantities is obtained  
by eliminating the bare quark mass in
computing the line of critical endpoints, 
\be \label{tline}
T^*(\mu_B)=T_c(m_c(\mu_B),\mu_B)=
T_c(m_c(0),0)\left(1-0.0060(4)\left(\frac{\mu_B}{T}\right)^2\right)\;.
\ee
While in this function we lose the information 
on the quark mass dependence, the curve relates only infrared quantities 
and describes a physical property of the QCD parameter space, cf.~\fig \ref{eline}.

\section{\label{dis} Discussion}

Let us now try to compare our results to those obtained by other groups.
The first Taylor coefficient in $m_c(\mu^2)$ for the three flavor theory
was also calculated by means of Taylor expanded reweighting \cite{alt1}.
In the form of \eq (\ref{result1}) their result for the coefficient is $67(19)$, compared
to ours of $0.82(36)$.
We observe that, continued to imaginary $\mu$, this
result violates the bound $c_1\leq 9$ derived in Sec.~\ref{exp}.
The only way to avoid this conclusion would be large $\op(\mu^4)$-effects, for which we
see no evidence. While at present we have no explanation for this rather drastic 
disagreement, we speculate that it is a statistics problem: 
the preliminary result of \cite{alt1} is
based on six thousand trajectories, and measurements for different $\mu$ are always
correlated in reweighting approaches. Considering the problems we mentioned in
Sec.~\ref{probs}, the similarly sobering findings of Ref.~\cite{nc}, and the scatter of our 
uncorrelated data in \fig \ref{B4_8}, this might account for the discrepancy. 

Eventually, we are of course interested in the 2+1 flavor theory with non-degenerate
masses. In this case the line of constant $B_4$ derived from the leading order
expression \eq (\ref{Bexp}) reads
\be
2(m_{u,d}-m_c(0))+(m_s-m_c(0))-A\mu^2=0\;.
\ee
However, a linear extrapolation in the quark mass to $am_s$ is 
most likely not valid.
Blindly substituting the bare quark masses of \cite{fk2} and our value for $A$, 
one would obtain a critical chemical potential $\mu_B\sim 3$ GeV. 
While this number is certainly meaningless, it seems nevertheless 
that our calculation would put the critical endpoint 
of the deconfinement line at considerably 
larger values of $\mu_B$ than those reported in \cite{fk2,alt1}.
To avoid extrapolations over large ranges, 2+1 flavor QCD with physical masses
requires additional calculations
in the light and heavier quark mass regimes and could
be quite different numerically.

Finally we would like to add one more comment concerning the difficulties
of distinguishing a first order phase transition from crossover, 
Sec.~\ref{probs}. Our discussion focused on the observable $B_4$, and one may ask
about its relevance for other methods of determining the endpoint, like finite size
scaling of susceptibilities or Lee-Yang zeroes. While other observables might well
have smaller statistical errors than $B_4$ when measured on the same number of
configurations, their relative behavior between first order and crossover regimes 
is nevertheless driven by the number of tunnelings, and therefore suffers from the same
slowness of the Monte Carlo history as our $B_4$ measurement, 
requiring similar statistics
in order to arrive at reliable results.

\section{Conclusions}

We have investigated the finite density phase diagram of three flavor QCD for
$\mu_B\lsi500$ MeV by means of
lattice simulations at imaginary chemical potential. Compared to previous studies
with $N_f=2,4$, 
we gathered much increased statistics allowing us to determine
the location of the critical line $T_c(m,\mu_B)$ through terms linear in the quark
mass and quartic in the chemical potential. 
The curvature of the critical line becomes more negative with increasing $N_f$.
Any mixing terms between quark mass and 
chemical potential are smaller than our present accuracy, 
rendering $T_c(m,\mu_B)/T_c(m,0)$ quark mass independent to a good approximation.

We have also studied the nature of the phase transition along the critical line,
and the location of its endpoint as a function of quark mass,
by studying the Binder cumulant as a function of quark mass and chemical potential.
We were able to compute the first coefficient of the critical quark mass 
$m_c(\mu^2)$ to 40\% accuracy. A constraint at the 90\% confidence level puts our result
at considerable odds with a preliminary result given by a Taylor expanded reweighting
technique \cite{alt1}, our critical endpoint being 
at larger $\mu_B$ for comparable quark masses. 
Our central results are given in \eqs (\ref{tc1}),(\ref{tc2}) and (\ref{result1}),(\ref{tline}).
While we have clearly demonstrated the feasibility of such a calculation, our results
exhibit the formidable difficulty of this task, whose unambiguous completion requires
computational resources beyond those presently available to us.
An extrapolation to the physical
$2+1$ flavor case requires additional simulations to account for the heavier
strange quark, and is envisaged for the future.

\section*{Acknowledgements}

Our simulations were performed on main frame platforms as well as PC clusters at
SCF Boston University, ETH Z\"urich, the University of Minnesota Supercomputer Institute, 
the RCNP at Osaka University, MIT 
and Jefferson Lab. We thank all those institutions for the provided resources, and in particular
F.~Berruto, A.~Nakamura, J.~Negele, A.~Pochinsky and C.~Rebbi for support in getting access to them.
We are also grateful to 
N.~Christ for providing us with Ref.~\cite{nc} prior to publication,
to C.~Schmidt for correspondence and to M.~Laine and K.~Rajagopal for useful comments 
on the manuscript.

\end{document}